\documentclass[sigconf]{acmart}

\usepackage{etoolbox}
\patchcmd{\quote}{\rightmargin}{\leftmargin 2em \rightmargin}{}{}

\usepackage{balance}
\usepackage{booktabs}
\usepackage{enumitem}
\usepackage{graphicx}
\usepackage{bm}
\usepackage{amsmath,amsfonts,amssymb}
\usepackage{subfigure}
\usepackage{multirow}
\usepackage{url}
\usepackage{color}
\usepackage{listings}

\newcommand{\ignore}[1]{}

\copyrightyear{2017}
\acmYear{2017}
\setcopyright{rightsretained}
\acmConference{SIGIR 2017 Workshop on Neural Information Retrieval (Neu-IR'17)}{}{\\August 7--11, 2017, Shinjuku, Tokyo, Japan}
\acmPrice{}
\acmDOI{}
\acmISBN{}
\acmPrice{}

\begin{document}

\title[]{An Exploration of Approaches to Integrating Neural Reranking Models in Multi-Stage Ranking Architectures}

\author{Zhucheng Tu, Matt Crane, Royal Sequiera, Junchen Zhang, and Jimmy Lin}
\affiliation{\vspace{0.1cm}
  \department{David R. Cheriton School of Computer Science}
  \institution{University of Waterloo, Ontario, Canada}
}
\email{{michael.tu,matt.crane,rdsequie,j345zhan,jimmylin}@uwaterloo.ca}

\begin{abstract}
We explore different approaches to integrating a simple convolutional
neural network (CNN) with the Lucene search engine in a multi-stage
ranking architecture. Our models are trained using the PyTorch deep
learning toolkit, which is implemented in C/C++ with a Python
frontend. One obvious integration strategy is to expose the neural network
directly as a service. For this, we use Apache Thrift, a software
framework for building scalable cross-language services.
In exploring alternative architectures,
we observe that once trained, the feedforward evaluation of neural
networks is quite straightforward. Therefore, we can extract the parameters of
a trained CNN from PyTorch and import the model into Java, taking
advantage of the Java Deeplearning4J library for feedforward evaluation. 
This has the advantage that the entire end-to-end system can be implemented in Java. As a
third approach, we can extract the neural network from PyTorch and
``compile'' it into a C++ program that exposes a Thrift service.
We evaluate these alternatives in terms of performance (latency and throughput) as well
as ease of integration.
Experiments show that feedforward evaluation of the convolutional
neural network is significantly slower in Java, while the performance of
the compiled C++ network does not consistently beat the PyTorch implementation.
\end{abstract}

\maketitle

\section{Introduction}

As an empirical discipline, information retrieval research requires
substantial software infrastructure to index and search large
collections. To address this challenge, many academic research groups
have built and shared open-source search engines with the
broader community---prominent examples include
Lemur/Indri~\cite{metzler2004combining,metzler2004indri} and
Terrier~\cite{ounis06terrier-osir,macdonald2012puppy}.
These systems, however, are relatively unknown beyond academic circles.
With the exception of a small number of companies (e.g., commercial web search engines),
the open-source Lucene search engine (and its derivatives
such as Solr and Elasticsearch) have become the
{\it de facto} platform for deploying search applications in industry.

There is a recent push in the information retrieval
community to adopt Lucene as the
research toolkit of choice. This would lead to a better alignment of
information retrieval research with the practice of building search
applications, hopefully leading to richer academic--industrial
collaborations, more efficient knowledge transfer of research
innovations, and greater reproducibility of research results. Recent
efforts include the
Lucene4IR\footnote{\url{https://sites.google.com/site/lucene4ir/home}}
workshop organized by Azzopardi et al.~\cite{azzopardilucene4ir}, the
Lucene for Information Access and Retrieval Research (LIARR)
Workshop~\cite{Azzopardi_etal_SIGIR2017} at SIGIR 2017, and the Anserini IR toolkit
built on top of Lucene~\cite{Yang_etal_SIGIR2017}.

Given the already substantial and growing interest in applying deep
learning to information retrieval~\cite{Mitra_Craswell_2017}, it would make
sense to integrate existing deep learning toolkits with open-source search
engines. In this paper, we explore different approaches to integrating
a simple convolutional neural network (CNN) with the Lucene search
engine, evaluating alternatives in terms of performance (latency and
throughput) as well as ease of integration.

Fundamentally, neural network models (and even more broadly,
learning-to-rank approaches)\ for a variety of information retrieval
tasks behave as rerankers in a multi-stage ranking
architecture~\cite{Matveeva_etal_SIGIR2006,Pedersen_SIGIR2010,Wang_etal_SIGIR2011,Tonellotto_etal_WSDM2013,Asadi_Lin_SIGIR2013,Culpepper_etal_ADCS2016,Mitra_Craswell_2017}.
Given a user query, systems typically begin with a
document retrieval stage, where a standard ranking function such as BM25 is
used to generate a set of candidates. These are then passed to one or
more rerankers to generate the final results. In this paper, we focus
on the question answering task, although the architecture is exactly
the same:\ we consider a standard pipeline architecture~\cite{Tellex_etal_SIGIR2003} where the
natural language question is first used as a query to retrieve a set
of candidate documents, which are then segmented into sentences and
rescored with a convolutional neural network (CNN). Viewed in isolation,
this final stage is commonly referred to as answer selection.

We explored three different approaches to integrating a CNN for answer
selection with Lucene in the context of an end-to-end question
answering system. The fundamental challenge we tackle
is that Lucene is implemented in Java, whereas many deep
learning toolkits (including PyTorch, the one we use) are written in C/C++
with a Python frontend. We desire a seamless yet high-performance way
to interface between components in different programming languages;
in this respect, our work shares similar goals as Pyndri~\cite{Pyndri} and Luandri~\cite{Luandri}.

Overall, our integration efforts proceeded along the following line of
reasoning:

\begin{itemize}[leftmargin=*]

\item Since we are using the PyTorch deep learning toolkit to train
  our convolutional neural networks, it makes sense to simply expose
  the network as a service. For this, we use Apache Thrift, a software
  framework for building scalable cross-language services.\footnote{\url{https://thrift.apache.org/}}

\item The complexities of deep learning toolkits lie mostly in
  training neural networks. Once trained, the feedforward evaluation
  of neural networks is quite straightforward and can be completely
  decoupled from backpropagation training. Therefore, we explored an
  approach where we extract the parameters of the trained CNN from PyTorch and
  imported the model into Java, using the Deeplearning4J library
  for feedforward evaluation. This
  has the advantage of language uniformity---the entire end-to-end
  system can be implemented in Java.

\item The ability to decouple neural network training from deployment
  introduces additional options for integration. We explored an
  approach where we directly ``compile'' our convolutional neural network into
  a C++ program that also exposes a Thrift service.

\end{itemize}

\noindent Experiments show that feedforward evaluation of the
convolutional neural network is significantly slower in Java, which
suggests that Deeplearning4J is not yet as mature and optimized as
other toolkits, at least when used directly ``out-of-the-box''.  Note
that PyTorch presents a Python frontend, but the backend takes
advantage of highly-optimized C libraries. The relative performance of
our C++ implementation and the PyTorch implementation is not
consistent across different processors and operating systems, and thus
it remains to be seen if our ``network compilation'' approach can
yield performance gains.

\section{Background}

In this paper, we explore a very simple multi-stage architecture for
question answering that consists of a document retrieval and answer
selection stage. An input natural language question is used as a
bag-of-words query to retrieve $h$ documents from the collection.
These documents are segmented into sentences, which are treated as
candidates that feed into the answer selection module.

\begin{figure}[t]
\centering\includegraphics[width=1.0\linewidth]{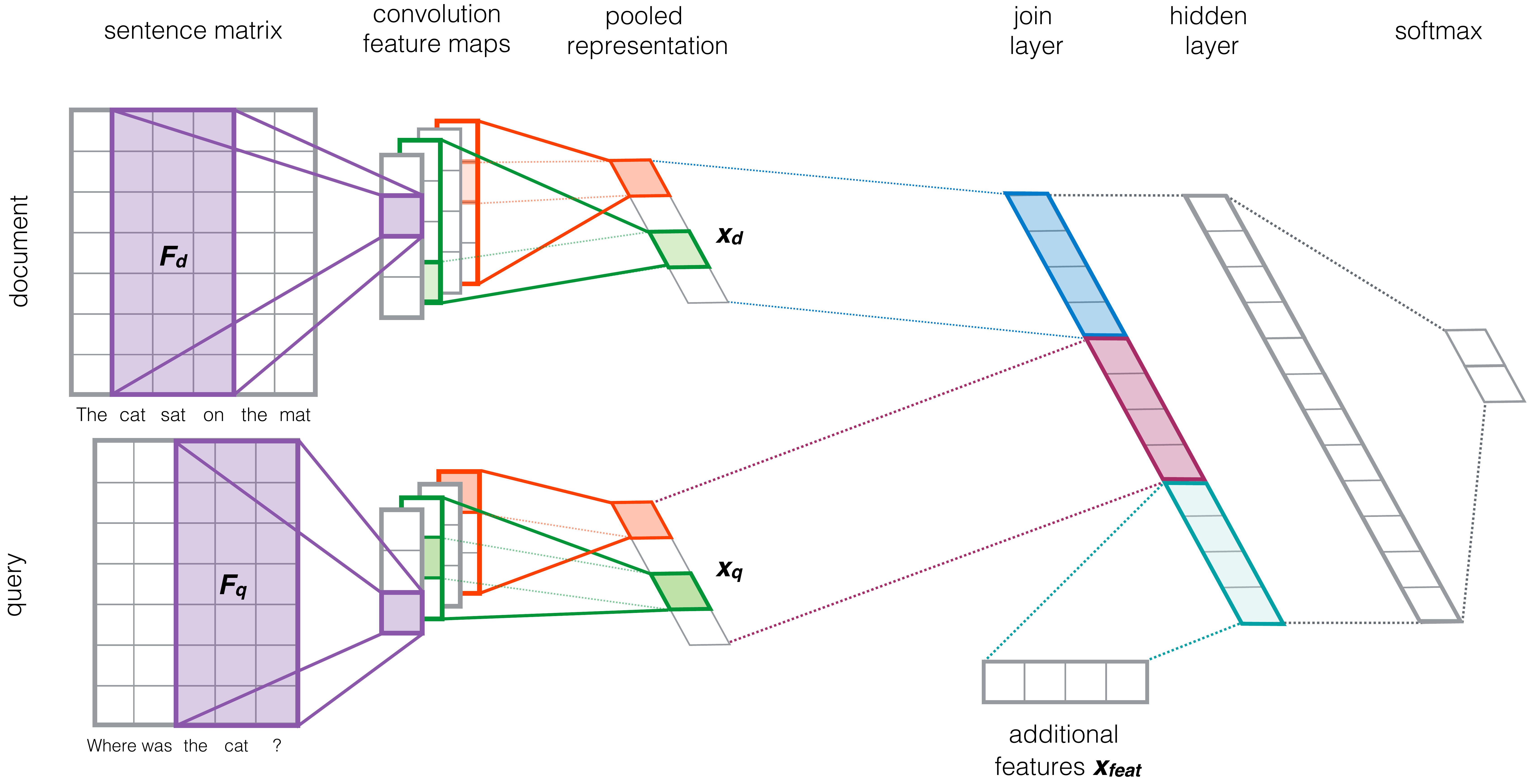}
\caption{The overall architecture of our convolutional neural network
  for answer selection.}
\label{figure:nn-architecture}
\end{figure}

Given a question $q$ and a candidate set of sentences $\{c_1, c_2,
\ldots c_n\}$, the answer selection task is to identify sentences that
contain the answer. For this task, we implemented the convolutional
neural network (CNN) shown in Figure~\ref{figure:nn-architecture},
which is a slightly simplified version of the model proposed by
Severyn and Moschitti~\cite{Severyn_Moschitti_SIGIR2015}.

We chose to work with this particular CNN for several reasons. It is a simple
model that delivers reproducible results with multiple
implementations~\cite{Rao_etal_SIGIR2017}. It is quick to train (even on CPUs), which supports fast
experimental iteration. Although its effectiveness in answer section is no longer
the state of the art, the model still provides a reasonable baseline.

Our model adopts a general ``Siamese'' structure~\cite{bromley-93}
with two subnetworks processing the question and candidate answers in
parallel. The input to each ``arm'' in the neural network is a
sequence of words $[w_1, w_2, \ldots w_{|S|}]$, each of which is
translated into its corresponding distributional vector (i.e., from a
word embedding), yielding a sentence matrix. Convolutional feature maps
are applied to this sentence matrix, followed by ReLU activation and
simple max-pooling, to arrive at a representation vector
$\textrm{x}_q$ for the question and $\textrm{x}_d$ for the ``document''
(i.e., candidate answer sentence).

At the join layer (see Figure~\ref{figure:nn-architecture}), all
intermediate representations are concatenated into a single vector:
\begin{equation}
\textrm{x}_{\textrm{join}} = [ \textrm{x}_q^T; \textrm{x}_d^T; \textrm{x}_{\textrm{feat}}^T ]
\end{equation}

\noindent The final component of the input vector at the join layer consists of
``extra features'' $\textrm{x}_{\textrm{feat}}$ derived from four word
overlap measures between the question and the candidate
sentence:\ word overlap and {\it idf}-weighted word overlap between
all words and only non-stopwords.

Our model is implemented using the PyTorch deep learning toolkit based
on a reproducibility study of Severyn and Moschitti's
model~\cite{Severyn_Moschitti_SIGIR2015} by Rao et
al.~\cite{Rao_etal_SIGIR2017} using the Torch deep learning
toolkit (in Lua). Our network configuration uses the best setting, as
determined by Rao et al.~via ablation analyses. Specifically, they
found that the bilinear similarity component actually decreases effectiveness,
and therefore is not included in our model.

For our experiments, the CNN was trained using the popular TrecQA
dataset, first introduced by Wang et al.~\cite{wang2007jeopardy} and
further elaborated by Yao et al.~\cite{yao2013answer}. Ultimately, the
data derive from the Question Answering Tracks from TREC
8--13~\cite{Voorhees_Tice_TREC8,Voorhees_Dang_TREC2005}.

\section{Methodology}

\subsection{PyTorch Thrift}

\begin{figure}[t]
\begin{lstlisting}[language=c]
namespace py qa

service QuestionAnswering {
    double getScore(1:string question, 2:string answer)
}
\end{lstlisting}
\vspace{-0.3cm}
\caption{Thrift IDL for a service that reranks question-answer pairs.}
\label{figure:thrift-idl}
\end{figure}

Given that our convolutional neural network for answer selection (as
described in the previous section) is implemented and trained in
PyTorch, the most obvious architecture for integrating disparate
components is to expose the neural network model as a service. This
answer selection service can then be called, for example, from a Java
client that integrates directly with Lucene. For building this
service, we take advantage of Apache Thrift.

Apache Thrift is a widely-deployed framework in industry for building
cross-language services. From an interface definition language (IDL),
Thrift can automatically generate server and client stubs for
remote procedure calls (RPC). In our
case, the server stub is in Python (to connect to PyTorch) while the
client stub is in Java (to connect to Lucene). These stubs handle
marshalling and unmarshalling parameters in a language-agnostic manner, message
transport, server protocols, connection management, load balancing, etc.

Our Thrift IDL for answer selection is
shown in Figure~\ref{figure:thrift-idl}. The namespace specifies the module that
the definition is declared in, as well as the language. The syntax for the function
declaration closely resembles a function declaration in C/C++ or Java, containing the
return type of the function as well as the name and type of each parameter.
The main difference is the addition of an integer parameter id, which is used for
identifying fields for schema evolution.

\begin{figure}[t]
\begin{lstlisting}[language=python]
class QuestionAnsweringHandler:
  ...

  def getScore(self, question, answer):
    xq = self.__make_input_matrix(question)
    xa = self.__make_input_matrix(answer)

    pred = self.model(xq, xa, self.default_ext_feats)
    pred = torch.exp(pred)
    return pred.data[0, 1]

class QAModel(nn.Module):

  def __init__(self, input_n_dim, filter_width,
          conv_filters=100, ...):
    ...
    self.conv_q = nn.Sequential(
      nn.Conv1d(input_n_dim, conv_channels,
            filter_width, padding=filter_width-1),
      nn.Tanh()
    )
    self.conv_a = ...
    self.combined_feature_vector =\
            nn.Linear(2*conv_channels, n_hidden)
    self.combined_features_activation = nn.Tanh()

  def forward(self, question, answer, ext_feats):
    q = self.conv_q.forward(question)
    q = F.max_pool1d(q, q.size()[2])
    q = q.view(-1, self.conv_channels)
    a = ...

    x = torch.cat([q, a, ext_feats], 1)
    x = self.combined_feature_vector.forward(x)
    x = self.combined_features_activation.forward(x)
    ...
    return x


\end{lstlisting}
\vspace{-0.3cm}
\caption{Snippet of the PyTorch implementation of answer selection
  with the Thrift service handler. Note the \texttt{getScore} method
  implements the Thrift interface and implicitly invokes the
  \texttt{forward} method of \texttt{QAModel}, the CNN in
  Figure~\ref{figure:nn-architecture}.}
\label{figure:python-impl}
\end{figure}

Thrift automatically generates ``service boilerplate'', and the
programmer is left to implement the service and client handler.
The service handler is essentially a wrapper that feeds the input
received by the server stub into the neural network model already implemented in
PyTorch, as shown in Figure~\ref{figure:python-impl}. The client directly calls
the corresponding method of the client stub generated by Thrift like an
ordinary function invocation---in our case, from Java.

\subsection{Deeplearning4J}

In our end-to-end question answering pipeline, only the answer
selection stage involves deep learning and thus depends on
PyTorch. However, the {\it training} of neural networks can be easily
decoupled from its deployment. Since the feedforward pass of neural
networks is relatively straightforward, we can use an existing
deep learning toolkit for training, and once the model is learned, we
can extract the model parameters and import the model into a different
toolkit (in a different programming language), or reimplement the 
feedforward evaluation directly in a language of our choice.

In our case, if we implement the feedforward evaluation in Java, then
our entire question answering pipeline can be captured in a single
programming language. We take advantage of a Java deep learning
toolkit called Deeplearning4J,\footnote{\url{https://deeplearning4j.org/}} built on
top of a Java numeric computation library called ND4J, to explore this
idea (see example code snippet in Figure~\ref{figure:java-impl}). Note
that although Deeplearning4J is a complete deep learning toolkit with
support for training neural networks, we only use it for forward
evaluation.  Since deep learning today remains dominated by
Python-based toolkits such as PyTorch, TensorFlow, and Keras, this
captures the workflow of most researchers and practitioners.

\begin{figure}[t]
\begin{lstlisting}[language=java]
MultiLayerConfiguration conf =
  new NeuralNetConfiguration.Builder()
    .list()
    .layer(0, new ConvolutionLayer.Builder(50, kernelW)
                .nIn(inChannels)
                .stride(stride, stride)
                .nOut(outChannels)
                .padding(0, padding)
                .activation(Activation.TANH)
                .build())
    .build();

model = new MultiLayerNetwork(conf);
model.output(batchedSentenceEmbedding, false);
\end{lstlisting}
\vspace{-0.3cm}
\caption{Snippet of the Deeplearning4J approach, showing how the convolution operation
is invoked.}
\label{figure:java-impl}
\end{figure}

To export model parameters, we need an interoperable serialization format for
exchanging data between Python and Java.
For this, we use Apache Avro, a widely-adopted data serialization framework.\footnote{\url{https://avro.apache.org/}}
Although Thrift can also be used for data serialization,
Avro's main advantage is its dynamic typing feature that avoids code generation. A script
written in Python loads the model trained by PyTorch, extracts the parameters,
parses the schema, and then writes the model parameters to an Avro file. The Avro schema
we use, expressed in JSON, is shown in Figure~\ref{figure:avro-schema}.

\begin{figure}[t]
\begin{lstlisting}[language=c]
{
  "type": "record",
  "name": "weights",
  "namespace": "CNN",
  "fields": [
    {
        "name" : "dimension",
        "type": {
          "type": "array",
          "items": "int"
        }
    },
    {
        "name" : "weights",
        "type": {
          "type": "array",
          "items": "double"
        }
    }
  ]
}
\end{lstlisting}
\vspace{-0.3cm}
\caption{Avro schema defining model weights.}
\label{figure:avro-schema}
\end{figure}

Note that weights for different types of parameters could have different
dimensions. For example, since multiple convolution filters are used to extract
different features from the sentence matrices, and each filter is a
2-dimensional tensor (matrix), the convolution filters together form a
3-dimensional tensor. However, the edge weights for the fully-connected layer
is only a 1-dimension tensor (vector). All weights are reshaped to one
dimension in the serialized storage, but are restored to their original
dimensions using saved dimension metadata. A Java program can read from the Avro
file, parse the schema, read the dimensions, reshape each weight matrix, and
finally convert them into the correct datatype for Deeplearning4J.

The advantage of the Deeplearning4J approach for reranking is language
uniformity, which allows the entire question-answering pipeline to be
tightly integrated into a single monolithic codebase---in contrast to
the microservices architecture that is fashionable today. There are
advantages and disadvantages to tightly-\ vs.\ loosely-coupled
architectures, but a broader discussion is beyond the scope of our work.

\subsection{C++ Thrift}

Once we separate the training of a neural network from its deployment,
we can in principle reimplement the feedforward pass in any language
we choose. A Java implementation (as described in the previous
section) yields language uniformity, which is a worthy goal from an
integration perspective. As an alternative, we explored the potential
for increased performance by ``compiling'' the neural network itself
into a standalone binary that exposes a Thrift service.

In more detail:\ We have implemented a
C++ conversion program that reads the same serialized model from Avro
(as described in the previous section). However, rather than construct the
network after reading the weights, the program instead generates another program with
the network already encoded---see representative snippet of code in
Figure~\ref{figure:cpp-impl}. This gives the maximum potential to benefit from
compiler optimizations as a number of the matrices are constants.
Since the conversion program is under our control, this technique can adapt to
arbitrary network architectures (although this is not the case in our
current proof-of-concept implementation). Additional, our compilation
approach means that any potential deployment requires only a single binary.

\begin{figure}[t]
\begin{lstlisting}[language=c++]
class QuestionAnsweringHandler
        : virtual public QuestionAnsweringIf {
  public:
  ... // Static matrix values
  double getScore(const std::string &question,
                  const std::string &answer) {
    // Convolution
    this->conv_result.resize(question_words.size());
    for (int i=0; i<this->q_conv_filters.size(); i++) {
      for (int k=0; k<question_words.size(); k++) {
        auto sub = submatrix(
          this->question_input, 0,
          k + COLUMN_PADDING,
          this->q_conv_filters[i].rows(),
          this->q_conv_filters[i].columns());

        auto cc = sub % this->q_conv_filters[i];
        float sum = 0.0;
        for (int j=0; j<cc.rows(); j++) {
          sum += 
            std::accumulate(cc.begin(j), cc.end(j), 0.0);
        }
        this->conv_result[k] = sum;
      }
      this->question_conv_map[i] =
          max(this->conv_result);
    }
    this->question_conv_map = tanh(
      this->question_conv_map +
      this->q_conv_biases);
    ...
    return score;
  }
};
\end{lstlisting}
\vspace{-0.3cm}
\caption{Snippet of the C++ generated program deployed behind the Thrift service,
showing the convolution computation for the question.}
\label{figure:cpp-impl}
\end{figure}

The generated program uses the Blaze math
library\footnote{\url{https://bitbucket.org/blaze-lib/blaze}}~\cite{Iglberger_etal_2012}
to perform all vector/matrix manipulations, which makes extensive use of
BLAS/LAPACK functions and SIMD instructions for maximum performance. Like the
PyTorch implementation, the generated program exposes a Thrift service. The
generation program and a simple Thrift client are publicly available on
GitHub.\footnote{\url{https://github.com/snapbug/coconut}}

\subsection{Alternative Approaches}
\label{section:alternative}

For completeness, it is worthwhile to discuss other obvious integration
approaches that we did not examine in this study.

In the quest for language uniformity, it would also be possible to
implement our entire CNN (including training) in Java using the
Deeplearning4J toolkit. We decided
not to pursue this route for two reasons:\ First, Deeplearning4J does
not appear to be as mature and widely used as other deep learning
toolkits such as Torch and TensorFlow. Second, we did not have an
implementation of our CNN in Deeplearning4J and lacked the resources
to port the model. We concede, however, that a thorough evaluation
should include this experimental condition.

There are alternative approaches to integrating Python and Java code
that we did not explore:\ PyLucene is a Python extension for accessing
Java Lucene, and Py4J is a bridge that allows a Python program to
dynamically access Java objects in a Java Virtual Machine. In such an
architecture, the entire question-answering pipeline would be
implemented in Python, calling Java for functionalities that involve
Lucene. Again, we did not have sufficient resources to explore this
approach to integration, but agree that such a condition should be
evaluated in future work.

\section{Results}

We compared the performance of our three proposed approaches in terms
of latency and throughput. All the experiments in this paper were
performed on two machines:

\begin{itemize}[leftmargin=*]

\item A desktop with an Intel Core i7-6800K CPU (6 physical cores, 12
  logical cores with hyperthreading) running Ubuntu 16.04.

\item An Apple MacBook Air laptop with an Intel Core i5-5250U CPU (2
  physical cores, 4 logical cores with hyperthreading) running
  macOS Sierra 10.12.6.

\end{itemize}

\noindent For the Python implementation, we used Python 3.6 and
PyTorch 0.1.12. For the Java implementation, we used Java 8 and
Deeplearning4J 0.8.0 on Oracle JVM 1.8.0. Deeplearning4J (which uses
the ND4J matrix library) uses OpenBLAS by default. Our C++
implementation uses Blaze 3.1 and was compiled with {\tt -O3 -DNDEBUG}
options using {\tt gcc} 5.4 on the desktop and Apple LLVM version
8.1.0 (clang-802.0.42) on the laptop.

Both PyTorch and Deeplearning4J use the {\tt OMP\_NUM\_THREADS}
environment variable to decide how many threads to use. We set
\texttt{OMP\_NUM\_THREADS=1} for all experiments. The C++
implementation is also single-threaded. In the relevant
configurations, Avro 1.8.2 and Thrift 0.10.0 are used. All
implementations ran on the CPU only. Performance measurements do not
include deserialization time of the model parameters and other startup
costs.

\subsection{Feedforward Evaluation}

We first examined the performance of the three approaches without
Thrift. For each implementation, after we loaded the model parameters
into memory, we iterated through the question--answer pairs from the
TrecQA raw-dev and raw-test splits and invoked the appropriate
function that computes the similarity score for each pair. We divided
the total number of pairs scored by the total elapsed time to arrive
at performance in terms of queries per second (QPS). Results are
summarized in Table~\ref{stats}. The calling program ran on a single
thread.

We observe that the performance of the Java Deeplearning4J
implementation is approximately 2--6$\times$ slower than the PyTorch
and C++ implementations, depending on the environment. In our initial
Java implementation, we used the ND4J matrix library directly to
implement convolutions in the simplest way possible---looping
over filters and convolving each filter with the sentence embedding
separately. The performance of this na\"ive implementation was two
orders of magnitude worse than that of Deeplearning4J, because
internally Deeplearning4J implements convolutions using the
\texttt{im2col} approach, which turns the problem into a matrix
multiplication and takes advantage of highly-optimized functions such
as GEMM.\footnote{\url{https://petewarden.com/2015/04/20/why-gemm-is-at-the-heart-of-deep-learning/}}
We spent a considerable amount of time investigating the performance
of our Java implementations (both using ND4J directly and
using Deeplearning4J). We were unable to further improve the
performance of the Deeplearning4J implementation; it seems that the
toolkit already exploits all the ``obvious'' optimization tricks.

Interestingly, the relative performance of the C++ and PyTorch
implementations is not consistent across our different test
environments, which is likely attributable to some combination of
hardware, operating system, and the software toolchain (e.g., {\tt
  gcc} vs.\ LLVM). We also note that the performance gap between the
Java implementation and the other two vary across the different
environments. Although performance differences are to be expected, we
currently lack a clear understanding of their sources.

Overall, these results suggest that Deeplearning4J and ND4J are less
mature than their counterparts in Python/C++ (i.e., PyTorch and numpy)
and ``out-of-the-box'' configurations may need further fine-tuning to
compete with the other approaches. The lack of a consistent
performance advantage in our C++ implementation suggests that the
Python frontend to PyTorch adds minimal performance overhead (since
the PyTorch backend also uses optimized C libraries). It remains to be
seen if further optimizations in our C++ implementation can translate
into consistently better performance.

\begin{table}[t]
\centering
\begin{tabular}{llr}
\toprule
Machine & Approach   & Throughput (QPS) \\
\midrule
Desktop & PyTorch &  1226.49 \\
& Deeplearning4J   & 530.4 \\
& C++ Generator  & 1235.50 \\
\midrule
Laptop & PyTorch &  828.57 \\
& Deeplearning4J   & 165.75 \\
& C++ Generator  & 1025.91 \\
\bottomrule
\end{tabular}
\vspace{0.2cm}
\caption{Feedforward evaluation performance of the CNN (without the
  Thrift service wrapper).}
\label{stats}
\end{table}

\subsection{Thrift Service}

Finally, we compared the performance of the PyTorch and C++
implementations behind Thrift. We did not examine wrapping
Deeplearning4J in a Thrift service since the advantage of the Java
implementation is the ability to directly integrate with Lucene. We
expect Thrift to introduce some overhead due to data
serialization/deserialization and network protocols, but such a design
enables components in an end-to-end question answering pipeline to be
built in different languages.

Typically, in a microservices architecture, the Thrift server and
clients would run on separate machines, but for the purposes of this
experiment, we run them both on the same machine. In this setup, we
use a Python Thrift client to make requests to both the Python Thrift
server as well as the C++ Thrift server. The Python client uses a
single thread to send requests---keeping the configuration as close to
the non-Thrift setting as possible. Both Thrift servers use
\texttt{TSimpleServer}, which runs a single thread, accepts one
connection at a time, and repeatedly processes requests from the
connection. Results are summarized in Table~\ref{e2e-stats}, with p50
and p99 denoting median and 99th percentile latency, respectively.

\begin{table}[t]
\centering
\begin{tabular}{llrrr}
  \toprule
&  & Throughput & \multicolumn{2}{c}{Latency (ms)} \\
  Machine & Approach  & (QPS) & p50 & p99 \\
\midrule
Desktop & PyTorch Thrift & 1150.86 & 0.83 & 1.72\\
& C++ Thrift   & 1000.40 & 1.00 & 1.59 \\
\midrule
Laptop & PyTorch Thrift & 774.28 & 1.20 & 2.51\\
& C++ Thrift   & 924.39 & 1.08 & 1.71 \\
\bottomrule
\end{tabular}
\vspace{0.2cm}
\caption{End-to-end performance of the Thrift service.}
\label{e2e-stats}
\end{table}

Comparing the results in Table~\ref{stats} with those in
Table~\ref{e2e-stats}, we are able to quantify the overhead introduced
by the Thrift service. On both machines, wrapping PyTorch with Thrift
added about 6--7\% overhead. Wrapping C++ with Thrift added around 10\% and
24\% overhead on the laptop and on the desktop, respectively. In this
particular case, it seems that the Python Thrift implementation is
more efficient than the C++ one.

\section{Conclusions}

In light of the recent push in the information retrieval community to
adopt Lucene as the research toolkit of choice and the growing
interest in applying deep learning to information retrieval problems,
this paper explores three ways to integrate a simple CNN with
Lucene. We can directly expose the PyTorch model with a Thrift
service, extract the model parameters and import the model into the
Java Deeplearning4J toolkit for direct Lucene integration, or
``compile'' the network into a standalone C++ program that exposes a
Thrift service.

All considered, the simplest approach of wrapping PyTorch in Thrift
seems at present to be the best option that combines performance and
ease of integration. The other two approaches have potential
advantages that remain currently unrealized. For Java, the performance
gap might shrink over time as Deeplearning4J becomes more mature. For
our compiled C++ approach, additional optimizations might yield
consistent performance gains to justify the additional
complexity. Finally, there are other approaches discussed in
Section~\ref{section:alternative} that we have not yet explored. The
best approach to integrating neural networks with the Lucene search
engine for information retrieval applications remains an open
question, but hopefully our preliminary explorations start to shed
some light on the relevant issues.

\section{Acknowledgments}

We'd like to thank the Deeplearning4J team for help with their toolkit.
This work was supported by the Natural Sciences and Engineering
Research Council (NSERC) of Canada, with additional contributions from
the U.S.\ National Science Foundation under CNS-1405688. Any findings,
conclusions, or recommendations express\-ed do not necessarily reflect
the views of the sponsors.




\end{document}